\documentclass[prd,aps,twocolumn,showpacs,superscriptaddress,amsmath,floatfix,amssymb]{revtex4}
\usepackage[colorlinks=false]{hyperref}
\usepackage{graphicx}

\newcommand{\beqn}{\begin{eqnarray}}
\newcommand{\eeqn}{\end{eqnarray}}
\newcommand{\eq}[1]{(\ref{#1})}
\newcommand{\lr}[1]{ \left( #1 \right) }

\newcommand{\vev}[1]{ \langle \, #1 \, \rangle }

\newcommand{\tr}{ {\rm Tr} \, }

\newcommand{\ket}[1]{ \, | #1 \rangle }
\newcommand{\bra}[1]{ \langle #1 | \, }
\newcommand{\sign}{ {\rm sign} \,  }
\newcommand{\logo}{\\ \vskip -15mm \leftline{\includegraphics[scale=0.3,clip=false]{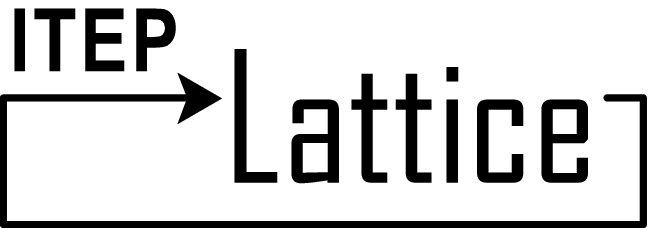}} \vskip 7mm}

\begin{document}
\sloppy

\title{Numerical evidence of chiral magnetic effect in lattice gauge theory\logo}

\author{P.V. Buividovich}
\affiliation{JIPNR ``Sosny'', National Academy of Science, Acad. Krasin str. 99, Minsk, 220109 Belarus}
\affiliation{ITEP, B. Cheremushkinskaya 25, Moscow, 117218 Russia}

\author{M.N. Chernodub}
\affiliation{LMPT, CNRS UMR 6083, F\'ed\'eration Denis Poisson, Universit\'e de Tours, 37200 France}
\affiliation{DMPA, University of Gent, Krijgslaan 281, S9, B-9000 Gent, Belgium}
\affiliation{ITEP, B. Cheremushkinskaya 25, Moscow, 117218 Russia}

\author{E.V. Luschevskaya}
\affiliation{ITEP, B. Cheremushkinskaya 25, Moscow, 117218 Russia}

\author{M.I. Polikarpov}
\affiliation{ITEP, B. Cheremushkinskaya 25, Moscow, 117218 Russia}

\date{August 13, 2009}
\begin{abstract}
The chiral magnetic effect is the generation of electric current of quarks along external magnetic field in
the background of topologically nontrivial gluon fields. There is a recent evidence that this effect is
observed by the STAR Collaboration in heavy ion collisions at RHIC. In our paper we study qualitative signatures of the
chiral magnetic effect using quenched lattice simulations. We find indications that the electric current
is indeed enhanced in the direction of the magnetic field both in equilibrium configurations of the quantum gluon
fields and in a smooth gluon background with nonzero topological charge. In the confinement phase
the magnetic field enhances the local fluctuations of both the electric charge and chiral charge densities.
In the deconfinement phase the effects of the magnetic field become smaller, possibly due to thermal screening.
Using a simple model of a fireball we obtain a good agreement between our data and experimental results
of STAR Collaboration.
\end{abstract}

\maketitle

\section{Introduction}

Properties of strongly interacting matter in hadron-scale magnetic fields have attracted a lot of attention recently.
The interest is motivated by feasibility to create such strong fields in ongoing heavy-ion experiments at Relativistic
Heavy Ion Collider (RHIC), and at future experiments at Facility for Antiproton and Ion Research at GSI, and ALICE
experiment at LHC. Noncentral heavy ion collisions may create magnetic fields due to relative motion of electrically
charged ions and the products of the collision.
For typical RHIC parameters the strength of the magnetic field at the center of a Au-Au collision may be estimated
to be of the order of $e B \sim \lr{10 - 100 \, \mbox{MeV}}^2$~\cite{Kharzeev:08:1} at first moments
($\tau \sim 1$\,fm/c) of the collision. The magnetic field created at RHIC is not only many orders of magnitude
stronger than any other experiment may reach, it also leaves behind even magnetized neutron stars,
magnetars, with their magnetic fields of the order of $10^{10}$\,Tesla (equivalent to $e B \approx 4\,\mbox{MeV}^2$).
Strong magnetic field can significantly modify the properties of nuclear matter affecting the spectrum
of the hadronic states~\cite{ref:Tiburzi}, shifting the position and changing the order of
the phase transition from hadronic matter to quark-gluon plasma~\cite{Agasian:Fraga:08:1:2:3}.

The magnetic fields lead to other unusual effects due to nontrivial
topological structure of
QCD~\cite{Kharzeev:08:1,Kharzeev:08:2,Kharzeev:98:1,Kharzeev:06:1}.
In this paper we concentrate on a particular realization of the
Chiral Magnetic Effect (CME), which generates a spatial separation
of positive and negative {\it electric} charges along the direction
of the {\it magnetic} field on nontrivial topological backgrounds of
gluons~\cite{Kharzeev:08:1,Kharzeev:08:2}. In the noncentral
collisions the magnetic field is perpendicular to the reaction
plane, therefore a charge separation can be observed in heavy-ion
collisions as a non-statistical asymmetry in the number of
positively and negatively charged particles emitted on different
sides of the reaction plane \cite{Voloshin:04:1,Voloshin:08:1}. This
effect was called ``event-by-event $\mathcal{P}$- and
$\mathcal{CP}$-violation'', because this asymmetry implies that the
difference of the numbers of quarks of different chiralities was
created in a particular event due to quantum fluctuations of the
topological charge. However, since there is no $\mathcal{CP}$ -
violation in strong interactions in the usual sense, this effect
vanishes after averaging over all events. The only way to observe it
experimentally is by studying $\mathcal{CP}$-even correlations of
reaction products \cite{Voloshin:04:1}. The magnetic field grows
with the impact parameter of the collision, therefore this asymmetry
should depend strongly on the centrality of the collision. There are
preliminary indications that this effect has been indeed observed by
the STAR collaboration at RHIC \cite{Voloshin:08:1,Caines:09:1}. It
should be stressed that the observed effect is
the indication that there are
fluctuations of chirality in QCD vacuum. A similar well-known effect
is the large mass of the $\eta'$ meson~\cite{Witten:79:2,
Veneziano:79:1}. However, the CME allows one to extract more
information on the fluctuations of chirality and topological charge,
since in this case there is an additional parameter -- the strength
of the magnetic field -- which can be varied in experiments.

The physical mechanism behind the CME is as follows~\cite{Kharzeev:08:1,Kharzeev:08:2}. At the energy scales of
the collisions, the light $u$ and $d$ quarks may be considered as approximately massless particles. The massless fermions
are characterized by right- or left-handed helicity (i.e., positive or negative projection of fermion's spin on the momentum,
respectively), so that the magnetic moment of (anti)quark is always collinear to its momentum. The strong enough external
magnetic field makes the magnetic moment to be parallel to the direction of the field so that the motion of the quarks
is essentially collinear to the magnetic field. The helicity is related to chirality: the quarks have the same
helicity and chirality while for the antiquarks the helicity is opposite to chirality. In the equilibrium ensembles the
total chirality is vanishing~\cite{McLerran:1990de}, so that the total electric current along the magnetic field is
zero (the left-handed and right-handed quarks move in opposite directions and compensate each other).
However, if there is a difference between the (anti)quarks with left- and right-handed chirality then the balance is
broken and the total electric current along the magnetic field becomes
nonzero~\cite{Nielsen:83:1, Metlitski:05:1, Kharzeev:98:1, Kharzeev:06:1, Kharzeev:08:1,Kharzeev:08:2}.

Thus, the CME arises due to the fluctuations in the numbers of left- and right-handed quarks in a vacuum or in
thermal state of non-Abelian gauge theories. In QCD the fluctuation of the chirality number are provided by
the topology-changing transitions between different vacua. Indeed, the gluon fields are
characterized by the topological winding number~\cite{tHooft:76:1,Zakharov:82:1}
\beqn
Q = \frac{g^2 }{16 \pi^2} \int d^4 x \, G_a^{\mu\nu}(x) {\widetilde G}^a_{\mu\nu}(x)\,, \qquad Q \in \mathbb{Z} \,,
\eeqn
where $G_a^{\mu\nu}$ is the gluon field strength tensor and the dual tensor is
${\widetilde G}_a^{\mu\nu} = (1/2)\, \varepsilon^{\mu\nu\alpha\beta}G_{a,\alpha\beta}$.
The configurations with $Q \neq 0$ change the chirality of the ensemble,
i.e. the difference between the number of left-handed quarks ($N_L$) and right-handed ($N_R$) quarks by an integer,
\beqn
\Delta(N_L - N_R) = N_f \, Q\,,
\label{eq:Delta}
\eeqn
where $N_f$ is the number of flavors. Therefore the electric current along the strong enough magnetic field
should be sensitive to the topological fluctuations of the gluon ensembles, providing us with possibility
to study in the direct way the topological structure of the QCD thermal states.

The CME is expected to work effectively in the chirally restored phase~\cite{Kharzeev:08:1,Kharzeev:08:2}:
the chiral condensate, which breaks the chiral symmetry, provides a direct coupling between left-handed quarks
to the right-handed ones thus leading to an efficient mechanism to suppress any asymmetry between them.
The suppression happens because the system prefers to decrease total chirality as the states with nonzero chirality
have higher energy due to the Fermi-principle~\cite{McLerran:1990de}. Therefore the CME should work effectively in
the high temperature case, where the chiral condensate is suppressed and the chirality fluctuations are strong.

While there are many phenomenological estimates of the strength of the CME in QCD, a detailed analytical calculation
is hardly possible due to strong nonperturbative effects. In this paper we report on our results of the numerical
investigation of the CME in quenched $SU\lr{2}$ lattice gauge theory both in high
and low temperature phases of the theory (Section~\ref{sec:simulations}). The chiral properties of the QCD, such as
chiral condensate~\cite{Buividovich:08:7} and the chiral magnetization~\cite{ref:magnetization}, in the
external magnetic field can reasonably be described by the quenched lattice theory. Our study involves
various approximations: we use quenched lattice gauge theory with two colors, the fermionic propagators are
truncated, and the theory is studied in a finite volume. Thus our results should only serve as a qualitative
indication of the existence of the CME, while the quantitative features may be accessible only in simulations
with dynamical quarks. Using our numerical results we nevertheless make a prediction for the charge asymmetry
and fit it to the experimental data in Section~\ref{sec:fit}. Our conclusions are summarized in the last Section.

\section{Lattice simulations}
\label{sec:simulations}

\subsection{Numerical setup}

We work in $SU\lr{2}$ lattice gauge theory with tadpole-improved Wilson-Symanzik
action (see, e.g., expression~(1) in Ref.~\cite{Luschevskaya:08:1}). In our zero-temperature
simulations we have used $14^4$ and $16^4$ lattices at lattice spacing $a = 0.103 \, \mbox{fm}$
and $a = 0.103 \, \mbox{fm}$, $a = 0.089 \, \mbox{fm}$, respectively.
The latter lattices were used to control the effects of finite volume and finite lattice
spacings, which turn out to be much less than the statistical errors.
For simulations at finite temperature we have used $16^3 \times 6$ lattices
with spacings $a = 0.128 \, \mbox{fm}$ and $a = 0.095 \, \mbox{fm}$, which correspond to the
temperatures $T = 256 \, \mbox{MeV} = 0.82 T_c$ and $T = 350 \, \mbox{MeV} = 1.12 T_c$, respectively.
The critical temperature in $SU(2)$ gauge theory is $T_c = 313.(3) \, \mbox{MeV}$~\cite{Bornyakov:07:1}.
As we explain below, the minimal nonzero value of magnetic field is rather large in our simulations,
$q B_{\mathrm{min}} \sim (350 \, \mbox{MeV})^2$).

In all current lattice simulations with background magnetic field only valence quarks
interact with electromagnetic field~\cite{Aubin:08:1,Lee:05:1,Buividovich:08:7,ref:Tiburzi},
because the inclusion of dynamical sea quarks into simulations make the problem significantly difficult.
On the other hand the characteristic properties of the CME may be studied in the quenched approximation,
in which the effects of the virtual quarks on the gluon fields  are neglected.

The quenched world has the same qualitative features as the unquenched
system: both approaches exhibit color confinement and chiral symmetry breaking. There is, however,
an important difference: due to the Crewther theorem~\cite{Crewther:1977ce}
fluctuations of the net topological charge in the theory with light virtual quarks should be drastically
suppressed compared to the quenched vacuum. However, as we have checked numerically, at least in the quenched approximation
the CME is mostly due to local fluctuations of the density of topological charge, rather than due to the global topological charge
of gauge field configurations. The number of colors is also inessential for the existence of the CME.
Thus we proceed with the quenched simulations of the $SU(2)$ gauge theory on the lattice.

For valence quarks we use Neuberger's overlap Dirac operator \cite{Neuberger:98:1}. The expectation
values of fermionic fields are first computed in terms of massive Dirac propagator in a fixed configuration of gauge
fields, according to the Wick theorem. For example, for the expectation value involving four fermionic
fields in  background of a fixed gauge field $A$, one has:
\begin{eqnarray}
\label{four_fermion_vev}
& & \vev{{\psi}^\dagger(x) O_{1} \psi(x) \, {\psi}^\dagger(x) O_{2} \psi(x)}_A \\
& &
= \tr\lr{\frac{1}{\mathcal{D} + m} \, O_{1}} \tr\lr{\frac{1}{\mathcal{D} + m} \, O_{2}}
\nonumber \\
& & \ - \tr\lr{\frac{1}{\mathcal{D} + m} \, O_{1} \, \frac{1}{\mathcal{D} + m} \, O_{2}}\,,
\nonumber
\end{eqnarray}
where $\mathcal{D}$ is the massless Dirac operator on the lattice,
\beqn
\mathcal{D} = \gamma^{\mu} \lr{\partial_{\mu} - i A_{\mu}}\,,
\eeqn
and $O_i$, $i=1,2$ are some operators in spinor and color space. These operators are treated as matrices in
spinor and color spaces. Consequently, the traces in Eq.~\eq{four_fermion_vev} are taken over
both spinor and color indices.

The overlap operator is a vital tool for investigation of the CME because this effect is realized for the chiral quarks, while the
overlap operator is the lattice Dirac operator which enjoys the chiral symmetry by the construction. On the contrary, if the quarks
are massive then the chiral magnetic effect is overshadowed by the flips of the quarks' spirality. Moreover, in the CME the
near-zero Euclidean eigenmodes play an essential role. The overlap operator gently treats low-lying eigenmodes making our results more
reliable. On the other hand, this operator is more complicated compared to other versions of the lattice
Dirac operators. As a result, the simulations are more power-consuming and our data has larger error bars.

The Dirac propagator $1/(\mathcal{D} + m)$ can be represented in terms of the eigenvalues $\lambda_k$
and the eigenmodes $\psi_k$ of the Dirac operator,
\beqn
\mathcal{D} \psi_k = i \lambda_k \psi_k\,,
\eeqn
as follows:
\begin{eqnarray}
\label{Dirac_propagator0}
\frac{1}{\mathcal{D} + m}\lr{x, y} = \sum \limits_{k} \frac{\psi_{k}\lr{x} \, \psi^\dagger_{k}\lr{y}}{i \lambda_{k} + m}\,.
\end{eqnarray}
Then Eq.~\eq{four_fermion_vev} can be reformulated as follows:
\begin{eqnarray}
\label{four_fermion_vev2}
& & \vev{\bar{\psi} O_{1} \psi \, \bar{\psi} O_{2} \psi}_A \nonumber \\
& &
= \sum \limits_{k,p} \frac{\bra{k} O_1 \ket{k} \bra{p} O_2 \ket{p} - \bra{p} O_1 \ket{k} \bra{k} O_2 \ket{p}}{(i \lambda_{k} + m)(i \lambda_{p} + m)}\,.
\end{eqnarray}

The result is then averaged over all configurations of the gauge fields.
This procedure is repeated for several values of the mass $m$ (we have used $m = 25\,\mbox{MeV}$, $m = 50\,\mbox{MeV}$, $m = 75\,\mbox{MeV}$), and the obtained dependence on $m$ is linearly extrapolated to zero to yield the expectation values in the chiral limit. In order to have a well-defined limit $m \rightarrow 0$, exact zero modes of the Dirac operator were not included in the Dirac propagator. The removal of zero modes is legal, since their contribution becomes insignificant in the limit $V \rightarrow \infty$, $m \rightarrow 0$ if one first tends the volume to infinity and then sets the mass to zero. The Dirac propagator is evaluated by inverting the massive Dirac operator in the subspace spanned by $M$ Dirac eigenvectors which correspond to $M$ nonzero Dirac eigenvalues with smallest absolute values. The truncated operator~\eq{Dirac_propagator0} reads as follows:
\begin{eqnarray}
\label{Dirac_propagator}
\frac{1}{\mathcal{D} + m}\lr{x, y} = \sum \limits_{|k| < M} \frac{\psi_{k}\lr{x} \, \psi^\dagger_{k}\lr{y}}{i \lambda_{k} + m}\,.
\end{eqnarray}
The value of $M$ is limited by the numerical procedure (\texttt{ARPACK} in our case) used to find the eigensystem of the Dirac operator.
We have found that the UV-finite quantities depend only weakly on $M$.

Uniform magnetic field with field strength $F_{12} = B$ is introduced into the Dirac operator by substituting
the $su\lr{2}$-valued vector potential $A_{\mu}$ with the $u\lr{2}$-valued one,
\begin{eqnarray}
\label{magnetic_field}
A_{\mu \, ij} & \rightarrow & A_{\mu \, ij} + A_{\mu}^{\lr{B}} \delta_{ij}\,,\\
A_{\mu}^{\lr{B}}\lr{x} & = & \frac{B}{2} \bigl(x_{1} \delta_{\mu, 2} - x_{2} \delta_{\mu, 1} \bigr)\,.
\nonumber
\end{eqnarray}
The expression (\ref{magnetic_field}) is valid in infinite space.
In order to combine it with periodic boundary conditions on the lattice we have introduced an additional $x$-dependent boundary twist for fermions, as explained in \cite{Wiese:08:1}. Since the total magnetic flux is quantized, on finite periodic space with the period $L$ the uniform magnetic field can take only discrete values:
\beqn
q B = \frac{2 \pi \, k}{L^{2}}\,, \qquad k \in \mathbb{Z}\,,
\label{eq:qB}
\eeqn
where $q$ is the quark charge. In our calculations we take $q = - e/3$, corresponding to the $d$-quark. Due to the quantization
condition~\eq{eq:qB} there is a minimal nonzero value of magnetic field, which is equal to
$\sqrt{q B} = 348 \, \mbox{MeV}$ in most our simulations (for $14^{4}$ lattice with lattice
spacing $a = 0.103 \, \mbox{fm}$ one has $L = 14 \cdot a = 1.44 \, \mbox{fm}$).

In our work the spatial tensor indices are $i, j = 1, 2, 3$ and the Euclidean time is labeled by the index $4$.
The magnetic field $F_{12} = B_{3} = B$ is uniform in our simulations, see Eq.~\eq{magnetic_field}.

\subsection{Observables and topology}

In order to characterize the CME quantitatively, we studied two basic quantities.
The first quantity is the local chirality
\beqn
\rho_{5}\lr{x} = \bar{\psi}\lr{x} \gamma_{5} \psi\lr{x}\,,
\label{eq:rho5:def}
\eeqn
which is the operator of the difference of the local densities of left- and right-handed quarks.

The second quantity is the electromagnetic current
\beqn
j_\mu\lr{x} = \bar{\psi}\lr{x} \gamma_\mu \psi\lr{x}\,.
\label{eq:j:def}
\eeqn
All other studied quantities are the powers and the products of \eq{eq:rho5:def} and \eq{eq:j:def}.
Note that in Euclidean space the spinor conjugation is just the complex
conjugation $\bar{\psi}_{\mathrm{Eucl}} = \psi^{\dag}$~\cite{Zakharov:82:1}.

According to the original idea of Refs.~\cite{Kharzeev:08:1,Kharzeev:08:2} the CME
appears in the presence of the topologically nontrivial configurations. The nonzero topological charge
is needed to destroy a balance between the left and right quark zero modes.  In general,
configurations of gauge fields may have different
values of the global topological charge. According to the Atyah-Singer theorem,
the topological charge of a gauge field configuration is given by a difference
of the right and left zero modes of the Dirac operator,
\beqn
N_L - N_R = Q
\label{eq:Q:lattice}
\eeqn
(this is essentially a Euclidean version of Eq.~\eq{eq:Delta} written for one species of fermions, $N_f = 1$). The crucial advantage of our numerical
procedure is that we are using the chirally symmetric Dirac operator which possess exact zero modes. The knowledge of the
zero modes for a given gluon field allows us to calculate the integer-valued topological charge of this configuration
using Eq.~\eq{eq:Q:lattice}.

Surprisingly, we have observed that all studied CME-related observables are independent of the global charge of the gluon fields. More
precisely, in the external magnetic field the effect of the nontrivial topology is smaller then the statistical error for any studied quantity
like \eq{eq:rho5:def} and \eq{eq:j:def}, their powers and correlations. One could incorrectly conclude that the electromagnetic properties
of the QCD vacuum in the external magnetic field have nothing to do with the topology of the gluon fields and, consequently, the CME is not
realized in the (thermal) equilibrium (we remind the reader that the CME is an interplay between the electromagnetism and the gluon topology).

Coming a bit in advance, we state that the real situation turns out to be more complicated. In fact, each gauge field configuration
contains quite strong {\it local} fluctuations of the topological charge. In the real equilibrium configurations of the gauge fields,
the density of the topological charge is localized on low-dimensional objects~\cite{ref:topological} which do not resemble smooth
structures like, for example, classical instantons. And, consequently, the effect of the global topology is completely overshadowed by strong local
fluctuations of the topological charge. Thus, below we do not classify the gauge field configurations by the topological charge, treating all
configurations on equal footing. On the contrary, our studies, reported in this paper, show that the CME does exist in equilibrium
and that the magnetic fields do affect the electromagnetic properties of vacuum at {\it local} concentrations of the topological charge.

We prefer to study correlation of the electromagnetic current~\eq{eq:j:def} with the
chiral density~\eq{eq:rho5:def}. The chiral density is calculated with the help of the low eigenmodes of
the Dirac operator in the background of the gauge field configuration, as described in the previous subsection.
In fact, the CME predicts the correlation of the electromagnetic
currents with the imbalanced density of the left and right quark zero modes. The exact zero modes of separate topological lumps
in a configuration of the gluon fields (one can think, for example, of separate instantons and anti-instantons) become near-zero modes because of the
interactions between the modes. This interaction breaks the zero-mode degeneracy and the former zero modes acquire tiny corrections.
Such near-zero modes carry information about unperturbed zero modes including the distribution of chirality and, eventually,
about the chiral magnetic effect.

In order to illustrate the insignificance of the exact Euclidean~\cite{foot2} zero modes
let us consider a configuration consisting of an instanton-antiinstanton pair. Assume that these objects are separated by a distance which is much larger compared to their sizes, and this configuration has no zero fermion modes since the total topological charge is zero. If one applies an external magnetic field to this configuration, the lumps of the topological charge will be the sources of the generation of the electric current.
At the instanton the electric current will be generated in one direction, while at the antiinstanton the current will start flowing in the opposite direction. The CME mechanism leads to the current generation because each of the (anti)instantons changes the chirality of quarks in the positive or negative direction, and the local imbalance of the chirality leads to a generation of the positive or negative current, respectively.

At the instanton-antiinstanton configuration the global electric current is zero, while locally each topological lump induces the electric currents exhibiting the CME. Therefore, we concentrate our attention on local correlators ignoring the global signatures of the CME like global currents and/or the Euclidean zero modes. Another argument in favor of local treatment of the CME is that in the Euclidean space the exact zero mode provides zero contribution to the electric currents and the electric charge density. This statement can readily be proven in the chiral basis of the Dirac matrices in which the $\gamma$-matrices are $2 \times 2$-diagonal structures. Nevertheless, the topological charge fluctuations do induce the electric currents because the in the presence on many positively and negatively charged lumps of topological charge  (one can think of,e.g., an ensemble of instantons and antiinstantons) the zero modes split into very-near-zero modes, which do remember about the local topology of the lumps. The low-lying modes
do contribute to the electric quark's currents and this fact is clearly seen in our simulations.

Thus, in all our calculations we study the nonzero fermionic eigenmodes and we ignore the exact zero modes.
Note that in large enough volumes the fermionic zero modes are suppressed and in the thermodynamic limit their contribution to certain physical observables -- for example, to the chiral condensate~\cite{Banks:1979yr} and to the chiral magnetization~\cite{ref:magnetization} -- is zero.
Indeed, as the volume increases the number of the exact zero modes growing as the square root of the total volume while the number of the
near-zero modes grows proportionally to the first power of the volume.

Finally, we mention that the space-time distribution and the dimensionality of the regions of space -- where the left/right currents are
imbalanced -- may be as complicated as the distribution of the topological charge itself. The
CME is expected to be realized due to the presence of the regions where the chiral density~\eq{eq:rho5:def} of the former zero-modes (and now near-zero modes) of quarks is enhanced. We can probe the mechanism of the CME by investigation of the mutual correlations between the chiral density~\eq{eq:rho5:def} and the electromagnetic current~\eq{eq:j:def}.

\subsection{Electric charge density: an illustration}

The basic feature of the CME is the spatial charge separation, which appears due to the presence of the nontrivial magnetic field $B$.
The separation is a result of the existence of the quarks' electric current induced in the direction of the magnetic field.
We call this current as the ``longitudinal'' electric current. According to our notations, the longitudinal direction is given by $\mu=3$
for $T\neq 0$ and it covers the whole $03$-plane for $T=0$. The transverse (with respect to the magnetic field) components of the electric
current belong to the $12$-plane.

In order to illustrate the CME we have first studied a spatial excess of the charge density due to the presence of the magnetic field,
\begin{eqnarray}
\label{eq:j:B}
j_0\lr{x;A,B} = \vev{j_0\lr{x}}_{A,B} - \vev{j_0\lr{x}}_{A,B=0}\,.
\end{eqnarray}
The subtraction in \eq{eq:j:B} removes all ultraviolet contributions leaving us with infrared nonperturbative
contributions originating due to the external magnetic field.

\begin{figure}[htb]
  \includegraphics[width=6cm,angle=0]{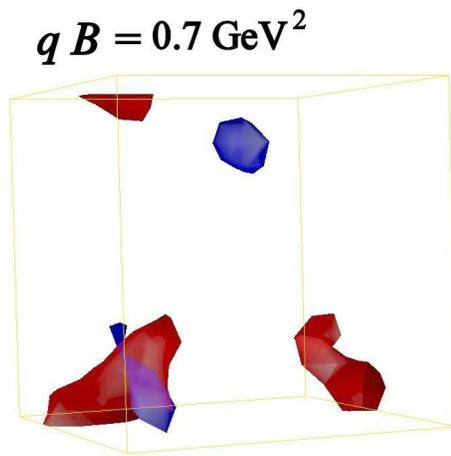}\\
  \caption{(color online). The excess of the density of the electric charge~\eq{eq:j:B} due to the applied magnetic field $qB = 0.7\,\mbox{GeV}^2$
  on $14^4$ lattice with the volume $(1.44 \, \mbox{fm})^4$. A typical three dimensional slice of a typical gauge field configuration
  is shown. The magnetic field is directed vertically from the bottom of the picture to the top. The red regions mark the excess of
  the positive charges while the blue regions correspond to the negative electric charge density.}
  \label{fig:j0H2}
\end{figure}

\begin{figure}[htb]
  \includegraphics[width=6cm,angle=0]{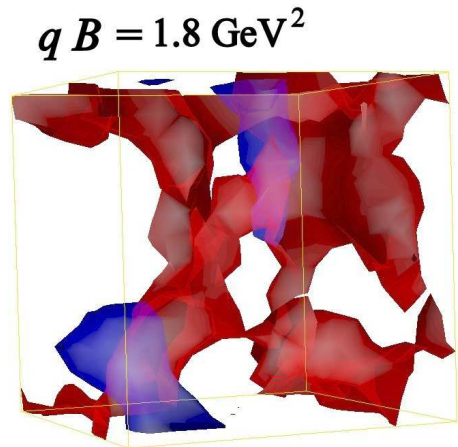}\\
  \caption{(color online). The same as in Figure~\eq{fig:j0H2} but for $qB = 1.8\,\mbox{GeV}^2$ and for the configuration of non-Abelian gauge field.}
  \label{fig:j0H5}
\end{figure}

Level surfaces of two typical charge density distributions in a fixed timeslice of a zero-temperature gauge
field configuration are shown in Figures~\ref{fig:j0H2} and \ref{fig:j0H5} for two values of the external magnetic field~$B$.

We can deduce three interesting qualitative features from Figures~\ref{fig:j0H2} and \ref{fig:j0H5}.

Firstly, we notice that the external magnetic field indeed induces a spatial excess of the electric charge density.
The stronger the field the larger the excess. This property is in agreement with our present understanding of the CME.

Secondly, the excess of the charge density is spatially extended along the direction of the magnetic field.
This feature can be understood in terms of the standard electrodynamics~\cite{ref:Landau:III}:
the quarks are localized in the transverse plane with respect to the field as they tend to occupy low Landau levels
(in Figures~\ref{fig:j0H2} and \ref{fig:j0H5} the transverse planes are horizontal). If there were no background gluon
field then the longitudinal quark trajectories are not affected by the field. In general quarks have a nonzero momentum
in the longitudinal direction, so that the trajectories of the quarks should be elongated strictly in the direction
of the magnetic field. The gluon background field provides a visible distortion to the quarks' motion. This feature is
observed in both figures.

Thirdly, Figures~\ref{fig:j0H2} and \ref{fig:j0H5} illustrate a conceptual difficulty associated with identifying the
electric dipole moment of the spatially separated charges in a finite volume. Due to the (periodic) boundary conditions
the electric dipole moment is not a universal notion as it depends on the choice of the reference frame. Consider, for
example, one dimensional space of the finite length $L$ with periodic boundary condition imposed. Two points separated by
the distance $r >0$ have the dipole moment $\mu_e = e r >0$.
However, if we change the direction of
our reference frame, one can count the distance between the two points as $L-r$. As a consequence, we get the different value for
the electric dipole moment $\mu_e = e (r-L) < 0$. The difference between the two definitions results in the large uncertainty of the
electric dipole moment, and
this fact makes the physical sense of the definition obscure. One can, of course, limit all distances on the lattice by $r \leqslant L/2$,
so that if $r < L/2$, then we define $\mu_e = e r$ while $r > L/2$ then $\mu_e = e (r-L)$. However, this definition is not satisfactory as well,
because small variations of the charge around the
most distant point $r \sim L/2$ lead to large variations of the electric dipole moment, $\mu_e \sim e L/2 \to - e L/2$. Another solution of this
problem could be to replace the infinite-volume definition $\mu_e = e r$ by its finite-volume counterpart, $\mu_e = e f(r,L)$,
where the function $f$: (i) obeys the symmetry $f(r,L) = f(L-r,L)$ and
(ii) reduces to $f \to r$ in the infinite volume limit, $L\to \infty$. However, this solution is not satisfactory either, because
the electric dipole moments in intermediate volumes would inevitably be dependent on the form of chosen function $f(r,L)$.

Thus, the electric dipole moment is an ambiguous notion in finite volume. Therefore, instead of the electric dipole moments
we study the electric currents. The induced currents represent a complimentary aspect of CME since
the emergence of the electric dipole moment is necessarily associated with an electric current
The advantage of the current-based approach is clear:
the definition of the electric current~\eq{eq:j:def} is a local notion which is not subjected -- at least
directly -- to the finite volume ambiguities.

\subsection{Chirality fluctuations and CME in cold matter}

Since the CME originates due to fluctuations of the local chirality, we first study the
effect of the external magnetic field on the expectation value of the square of the
local chirality~\eq{eq:rho5:def}:
\begin{eqnarray}
\label{chirality_sq_def}
\vev{ \rho_{5}^{2} }_{IR}\lr{B, T} = \frac{1}{V} \int \limits_{V} d^{4}x  \vev{\rho_{5}\lr{x} \rho_{5}\lr{x}}_{B, T}
\nonumber \\ -
\frac{1}{V} \int \limits_{V} d^{4}x \vev{ \rho_{5}\lr{x} \rho_{5}\lr{x}}_{B = 0, \, T = 0}\,.
\end{eqnarray}
Here $V$ is the total volume of the lattice, and $\vev{\ldots}_{B, T}$ means expectation value with respect to (w.r.t.)
the thermal state at temperature $T$ with background magnetic field $B$. Subtraction of the expectation value at zero temperature
($T = 0$) and in the absence of the external field ($B = 0$) removes all ultraviolet divergences providing us with a physical
quantity which should be
insensitive to the ultraviolet cutoff. In fact, the quantity~\eq{chirality_sq_def} provides us with the strength
of the chirality fluctuations because in the equilibrium QCD the average of the chirality is zero, $\langle \rho_5 \rangle = 0$.
We denote the result of the subtraction by the subscript ``IR'' in $\vev{\ldots}_{IR}$ stressing
the importance of the nonperturbative infrared (IR) contributions. In all our expectation values the
corresponding zero-temperature and zero-field contributions are always subtracted as it is done in (\ref{chirality_sq_def}).
Note that the coordinates of all operators entering our observables, including Eq.~\eq{chirality_sq_def}, are taken at one point $x$.

\begin{figure}[htb]
  \includegraphics[width=6cm, angle=-90]{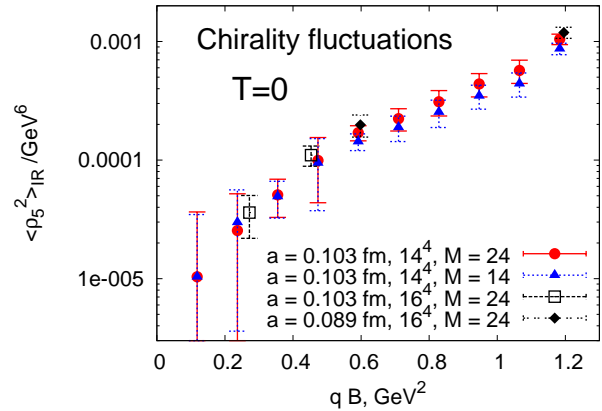}
\caption{(color online). The expectation value of the chirality squared~\eq{chirality_sq_def}, $\rho^2_5$, vs. magnetic field at different lattice
parameters (the lattice spacing $a$ and the lattice volume $L^4$), and for different number $M$ of Dirac eigenmodes which were taken
to determine the truncated Dirac propagator~\eq{Dirac_propagator} used in \eq{four_fermion_vev} to calculate~\eq{chirality_sq_def}.}
  \label{fig:rho52}
\end{figure}

Due to the subtraction, the expression~\eq{chirality_sq_def} gives us the influence of the magnetic field
and/or finite temperature on the strength of the chirality fluctuations. We plot this quantity in Figure~\ref{fig:rho52}
for a few values of the lattice spacing $a$, lattice volumes $L^4$ and the number of low eigenmodes $M$
which are used to calculate the Dirac propagator~\eq{Dirac_propagator}. We found that the variations of all these parameters change
the expectation values by less than 10\%, thus confirming good scaling of our results.

The CME appears due to the local topology of the gauge field configurations. The topological content of the gauge field
configurations is reflected in the structure of the exactly-zero and near-zero eigenmodes. The exactly zero eigenmodes are not significant
in the Euclidean space as we have discussed above. However, the near-zero modes play an important role in the CME, and they also come with
larger weights in the propagator~\eq{Dirac_propagator}. Thus, the consideration of only low-lying zero modes is justified. For our particular observables, the errors associated with our truncation scheme are smaller than the typical statistical errors.
However, we should make a cautionary remark that as the volume of the lattice increases, the density of low-lying eigenmodes grows, and one
should take into account more and more of them.

Notice that most of Figures in our paper are plotted in the logarithmic scale, and that due to dimensional reasons the data for chirality fluctuations
and, later, for current fluctuations are always expressed in GeV${}^6$. Due to the high (sixth) power the actual values of our quantities look quite small, but this fact in no way indicates that the physical effects are small. For example, in Figure~\ref{fig:rho52} at $qB = 1.5 \,\mbox{GeV}^2$
the value of the chirality fluctuation is $\langle \rho_5^2 \rangle = 10^{-4}\,\mbox{GeV}^6$. This number corresponds
to the large quantity $\langle \rho_5^2 \rangle \approx (215\,\mbox{MeV})^6$ if the magnitude is expressed in terms of MeV.

\begin{figure}
  \includegraphics[width=6cm, angle=-90]{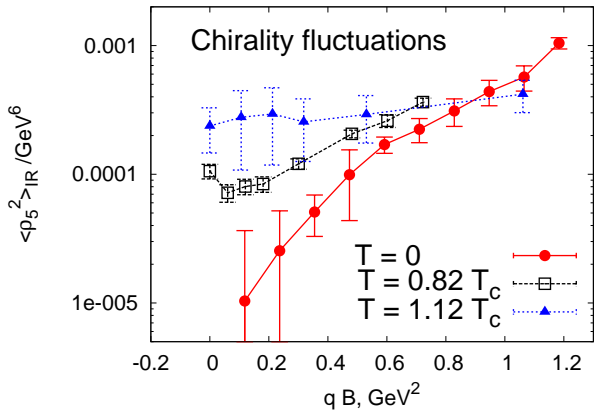}
  \caption{(color online). The expectation values of the chirality squared~\eq{chirality_sq_def} vs. the magnetic field at three different temperatures.}
  \label{fig:rho52_vs_H}
\end{figure}
In Figure~\ref{fig:rho52_vs_H} we show the dependence of the expectation value $\vev{ \rho_{5}^{2} }_{IR}\lr{B, T}$ on the magnetic field at three
temperatures: at zero temperature, at finite temperatures in the confinement phase ($T = 0.82 \, T_c$) and
in the deconfinement phase just above the phase transition, $T=1.12\, T_c$.
At zero temperature the fluctuations of the chirality quickly grow with magnetic field. As the temperature increases the grows rate gets smaller
and in the deconfinement phase the chirality fluctuation are almost insensitive to the value of the magnetic field.

In the absence of the external magnetic field, $B=0$, the rise of temperature leads, according to Figure~\ref{fig:rho52_vs_H},
to the increase of the chirality fluctuations.  However, sufficiently strong magnetic fields
can also induce the enhancement effect comparable to the effect of the thermal fluctuations.
For example, the chirality fluctuations at zero temperature in the magnetic field of the strength $q B \sim 2 \, \mbox{GeV}^2$
has the same value as the fluctuations of chirality in the absence of the external magnetic field at temperature $T = 1.12\, T_c$:
\beqn
\langle \rho_5^2(x)\rangle_{T=0,\ B = 2 \, {\mathrm{GeV}}^2} \approx \langle \rho_5^2(x)\rangle_{T=1.12 \, T_c,\ B = 0}\,.
\eeqn

It is known that the temperature suppresses the fluctuations of the {\it global} topological charge~\cite{Pisarski:80:1},
\beqn
\chi_Q = \frac{1}{V} \langle Q^2 \rangle\,.
\label{eq:chi:Q}
\eeqn
Lattice studies show that  in Yang-Mills theory the fluctuations are almost independent on the temperature in the confinement phase,
$$\chi_Q(T=T_c) \approx \chi_Q(T = 0)\,,$$
while as temperature exceeds the critical point, $T>T_c$, the fluctuations vanish quickly~\cite{Alles:97:1}.
For example, in SU(2) Yang-Mills theory just above the phase transition, $T = 1.1 \, T_c$, the value of the fluctuations
is more than twice smaller compared to the value in the confinement phase~\cite{Alles:97:1},
$$\chi_Q(T=1.1\,T_c) \approx 0.5\, \chi_Q(T = T_c)\,.$$
Moreover, it is clear that in quenched theory the topological charge of the gluon fields is not affected by the external magnetic field,
because the coupling between gluons and electromagnetism is provided by the dynamical quarks which are absent in the quenched theory.
These observations suggest that the {\it local} fluctuations of the chirality~\eq{chirality_sq_def} and
the fluctuations of the {\it global} topological charge~\eq{eq:chi:Q} are not related directly.
It is more likely that at high temperatures chirality-changing transitions occur due to local processes involving sphaleron-like
configurations~\cite{McLerran:91:1}. At zero temperature one might explain chirality changes as follows: in a strong magnetic field
quark undergoes one-dimensional motion along the magnetic field. Quark chirality can be efficiently changed in local one-dimensional
scattering processes on the fluctuations of gauge fields, where the spin is not affected but the momentum changes its sign.
In the deconfinement phase this mechanism does not work because the influence of the gluon fluctuations on the quark's motion is lower
in this phase (the quark confinement is lost, for example) and one can suggest that the typical scattering does not reflect the momenta.

Concluding this subsection, we notice that the CME crucially depends on the imbalance
between the left and right chiral modes. The local strength of the chiral imbalance
is characterized by the value of the chirality fluctuations, $\langle \rho_5^2 \rangle$,
which determines the rate of chirality-changing transitions. Since the external magnetic
field strongly enhances the chirality fluctuations (they become comparable with the
fluctuations in the deconfinement phase) so that the Chiral Magnetic Effect -- if we
imagine for a moment that the property of the color confinement were either absent or
inessential due to density effects -- could be observed in cold nuclear matter as well.

\subsection{Fluctuations of electromagnetic current}

We checked that the expectation value of all components of the electromagnetic current itself~\eq{eq:j:B} is zero within the error bars.
This fact is not unexpected because the magnetic field passes through the regions of space-time characterized
by positive and negative chiralities~\eq{eq:rho5:def}. According to the CME, in regions with opposite chiralities the magnetic
field induces the electric currents in opposite directions with respect to the magnetic field.  Thus, globally, the
longitudinal electric currents should cancel each other. As for the transverse (with respect to the magnetic field)
spatial directions, the electric currents should cancel each other because of circular motion on (perturbed) Landau levels.
Finally, the average timelike current is zero because of the Gauss theorem (the total electric charge in a closed volume with
periodic boundary conditions should be zero). These statements are valid both at zero and at non-zero temperatures.

Below we study the dispersion of the electromagnetic current
\begin{eqnarray}
\label{eq:j2:def}
& & \hskip -10mm
\vev{j_\mu^{2} }_{IR}\lr{B, T} = \frac{1}{V} \int \limits_{V} d^{4}x  \vev{j^2_\mu\lr{x}}_{B, T}
\nonumber \\
& & \hskip -8mm
- \frac{1}{V} \int \limits_{V} d^{4}x \vev{j_\mu^2\lr{x}}_{B = 0, \, T = 0}
\qquad \mbox{[no sum over $\mu$]}\,,
\end{eqnarray}
instead of the expectation value of the electric current.
Note that here and below no sum over the Lorenz index $\mu$ is assumed in all our expressions involving $j_\mu^2$.

\begin{figure}
  \includegraphics[width=6cm, angle=-90]{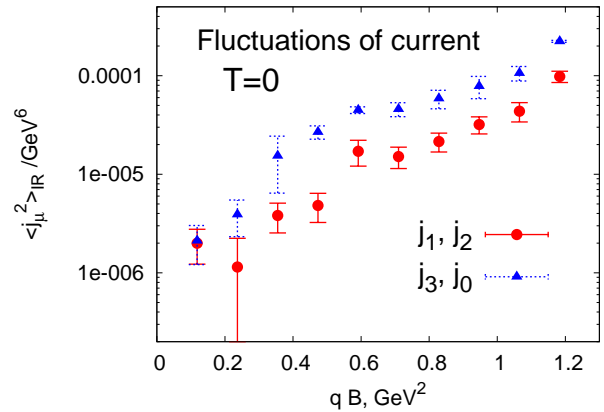}
  \caption{(color online). The expectation value of the squares of the transverse ($j_1$, $j_2$) and longitudinal ($j_3$, $j_0$) components
  of the vector current~\eq{eq:j:def} vs. the applied magnetic field at zero temperature.}
  \label{fig:j2vsH:T0}
\end{figure}

Following study of the fluctuations of chirality, we first check that the internal lattice parameters do not
influence the final results in Figure~\ref{fig:rho52}.
\begin{figure}
  \includegraphics[width=6cm, angle=-90]{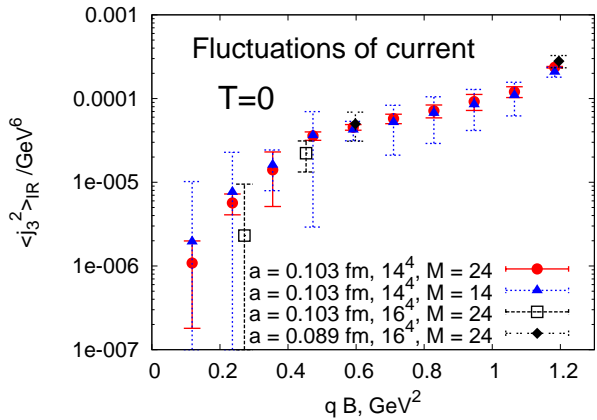}\\
  \caption{(color online). The same as in Figure~\ref{fig:rho52} but for the fluctuations~\eq{eq:j2:def}
  of the electromagnetic current $j_3$~\eq{eq:j:def}.}
  \label{fig:j2vsH}
\end{figure}
Then in Figure~\ref{fig:j2vsH:T0} we plot the expectation values of the fluctuations of each component of the electric current at zero temperature.
At $T=0$ the rotational symmetry is broken only by the electromagnetic field strength tensor $F_{\mu\nu}$ with $F_{12} \neq 0$, so that
the components of the electric currents can be grouped into transverse and longitudinal sets, respectively,
\beqn
\vev{j_{1}^{2}} = \vev{j_{2}^{2}}\,, \quad \vev{j_{0}^{2}} = \vev{j_{3}^{2}}\qquad (T=0)\,.
\eeqn
At weak magnetic fields all components have the fluctuations of the same order. However, at stronger fields the fluctuations of the longitudinal
currents ($j_3$ and $j_0$) are stronger compared to the fluctuations the transverse currents ($j_1$ and $j_2$). Both transverse and longitudinal components grow with magnetic field, but the longitudinal components grow faster. Thus we see that at zero temperature the magnetic field enhances
both the current $j_{3}$ in the direction of the magnetic field and the charge density $j_{0}$, which is a clear signature of the charge
separation associated with the CME. The enhancement of the transverse components $j_1$ and $j_2$ in magnetic field is natural since the
transverse momentum of a particle occupying a Landau level grows with the increase of the magnetic field.

At nonzero temperature the (discrete) rotational symmetry is broken not only by the electromagnetic field, $F_{12} \neq 0$, but also by
the compactified temperature direction $\mu=0$. Therefore the charge fluctuations ($\mu=0$) loose the similarity with longitudinal
($\mu=3$) current fluctuations, while the fluctuations of the transverse ($\mu=1,2$) currents are still degenerate,
\beqn
\vev{j_{1}^{2}} = \vev{j_{2}^{2}}\,, \quad \vev{j_{0}^{2}} \neq \vev{j_{3}^{2}}\qquad (T \neq 0)\,.
\label{eq:different}
\eeqn

\begin{figure}
  \includegraphics[width=6cm, angle=-90]{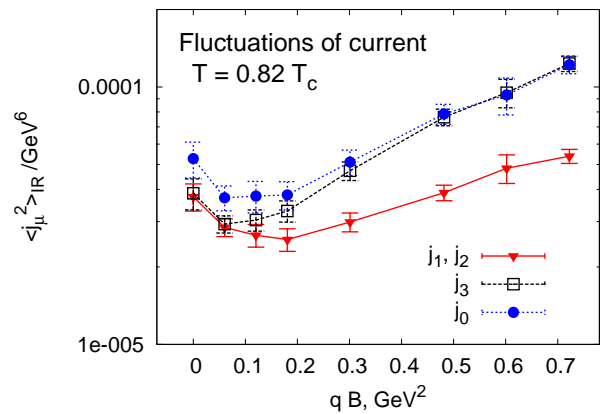}
  \caption{(color online). The same as in Figure~\ref{fig:j2vsH:T0} but for $T=0.82\, T_c$. The longitudinal components $j_3$ and $j_0$ differ due to nonzero temperature~\eq{eq:different}.}
  \label{fig:j2vsH:T082}
\end{figure}
In Figure~\ref{fig:j2vsH:T082} we plot the fluctuations of all components of the electric currents at finite temperature $T=0.82 \, T_c$,
at which the system is still in the confinement phase. One can see that at zero magnetic field the currents fluctuate strongly due to
thermal fluctuations.
All components of the electric currents fluctuate approximately with the same strength. As the magnetic field grows
the fluctuations of all components drop a bit, and then they start to grow steadily. The grows rate is noticeably smaller
compared to the slope observed at zero temperature, Figure~\ref{fig:j2vsH:T0}. On the other hand, there are similarities with the zero-temperature case: (i) the fluctuations of the charge density and of the longitudinal (with respect to the magnetic field) current have equal magnitudes;
and (ii) the longitudinal currents fluctuate stronger compared to the transverse currents.

\begin{figure}
  \includegraphics[width=6cm, angle=-90]{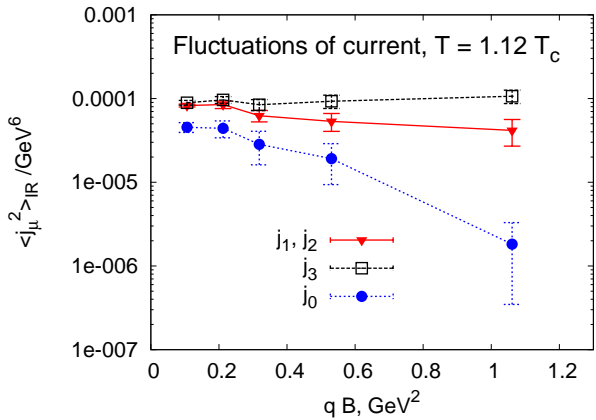}
  \caption{(color online). The same as in Figure~\ref{fig:j2vsH:T082} but for $T=1.12\, T_c$.}
  \label{fig:j2vsH:T112}
\end{figure}
The situation turns to be quite different in the deconfinement phase, Figure~\ref{fig:j2vsH:T112}.
First, the fluctuations of the longitudinal current are insensitive to the strength of the magnetic
field. Second, both the charge and transverse current fluctuations are decreasing functions of the
field strength. The drop in the electric charge fluctuations is especially noticeable. This feature
can be explained by an effect of the thermal Debye screening~\cite{foot1}.

\begin{figure}
  \includegraphics[width=6cm, angle=-90]{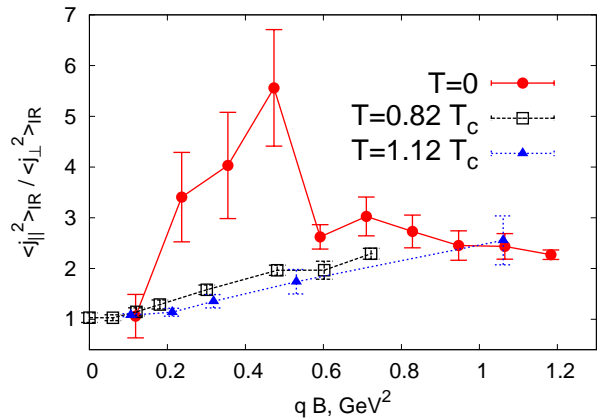}
  \caption{(color online). The ratio of the longitudinal current to the transverse one vs. the strength of the magnetic field at various temperatures.}
  \label{fig:ratio}
\end{figure}

In Figure~\ref{fig:ratio} we show the ratio of the longitudinal component of the current to its transverse component.
At zero temperature the current is substantially enhanced at $q B \sim 1.5 \,\mbox{GeV}^2$, while at relatively weak fields, $q B \sim 0.3 \,\mbox{GeV}^2$, the enhancement of the current is not seen.

At nonzero temperature the relative enhancement is weaker. The ratios at $T=0.82\, T_c$ and at $T=1.12\, T_c$ have similar properties
which are in a qualitative agreement with the CME features: at low magnetic fields the longitudinal and transverse directions are equivalent and the increase of the magnetic field gives a bit stronger preference for the longitudinal direction with respect to the transverse one.

We have already mentioned the reasons why the CME is expected to exist in the high temperature phase~\cite{Kharzeev:08:1}, where the confinement is absent and the chiral symmetry is restored. The reasons are quite simple: the presence of the chiral condensate destroys imbalance between the right and left chiral modes while the confinement does not allow the spatial separation of the quarks at the distances larger than the confinement distance. This argument seems to be independent on the inclusion of the dynamical (virtual) quarks into the consideration, so that it should be valid, in lattice language, both in quenched and in unquenched theories.

We have found that in the quenched approximation the qualitative signatures of the chiral magnetic effect exist both in the confinement and in the deconfinement phases. In all our simulations, at zero temperature and at finite temperature above and below the critical point, we see that the longitudinal component of the electric (vector) current dominates over the transverse component. However, in the confinement phase the electric current rises as the function of the external magnetic field while in the deconfinement phase the value of the current is almost insensitive to the strength of the magnetic field.

\subsection{Chirality-current correlations}

According to the CME the longitudinal electric current may locally be induced by the imbalance of chirality.
Therefore it is natural to expect that the local fluctuations of the longitudinal electric currents are correlated
with the chirality fluctuations. However, the correlator of the chirality and the electric current $\vev{\rho_{5} \, j_{\mu}}$
vanishes, since in isotropic homogeneous space it is impossible to
construct an axial vector from an antisymmetric tensor of magnetic
field strength.
Therefore, we have considered the correlation between the squares of the induced current and the chirality:
\begin{eqnarray}
\label{corrr52j2_def}
c\lr{ \rho_{5}^{2}, j_{\mu}^{2} } = \frac{ \vev{\rho_{5}^{2} \, j_{\mu}^{2}} -
\vev{\rho_{5}^{2}}\vev{j_{\mu}^{2}} }{ \vev{\rho_{5}^2} \vev{j_{\mu}^2}}\,.
\end{eqnarray}

\begin{figure}
  \includegraphics[width=6cm, angle=-90]{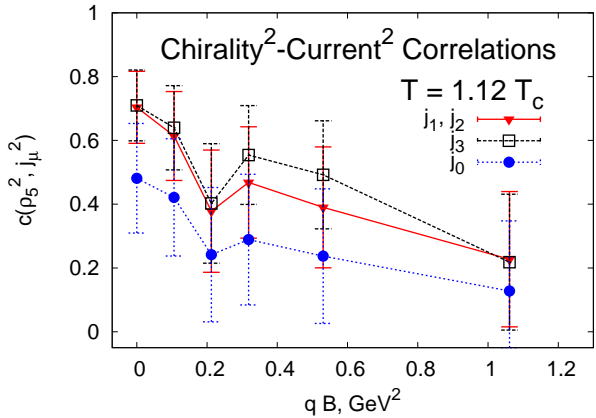}
  \caption{(color online). The correlation between squared chirality and squared electric current~\eq{corrr52j2_def}.}
  \label{fig:cr5_vs_H}
\end{figure}

It turns out that in the confinement phase the corresponding correlation functions
for all components of the electric currents vanish within statistical uncertainty,
\beqn
c\lr{ \rho_{5}^{2}, j_{\mu}^{2} } \simeq 0 \qquad T \lesssim T_c\,.
\eeqn
In the deconfinement phase the correlation function~\eq{corrr52j2_def} is not zero, Figure~\ref{fig:cr5_vs_H}, while
the correlation decreases with the increase of the magnetic field. All these facts suggest that the
enhancement of the fluctuations of current in the direction of the magnetic field is not
locally correlated with the chirality fluctuations.

We would like to emphasize, that the absence of the local correlation does not in general mean the independence of the longitudinal currents
on the chiral density. The longitudinal current can be caused by the chirality fluctuations, and, at the same time, the current itself may be
not localized in the local regions of the spacetime where the chiral density is enhanced. Although this statement sounds puzzling, similar
effects are already known in the condensed matter physics. For example, the well-known Anderson localization~\cite{ref:Anderson} of electron inside
a semiconductor is caused by the presence of impurities. However, the spatial regions, in which the electron wavefunctions are localized, do
not coincide with the positions of the impurities themselves.

\subsection{Clean case: instanton-like configuration}

\begin{figure}
  \includegraphics[width=6cm,angle=-90]{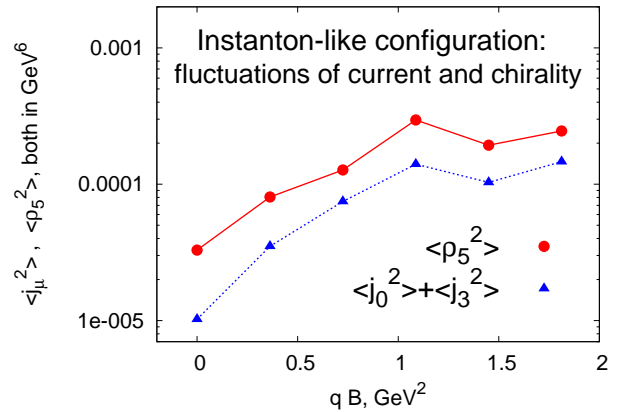}\\
  \caption{(color online). The fluctuations of the chirality~\eq{chirality_sq_def} and the fluctuations of the longitudinal electric current~\eq{eq:j2:def} at an instanton-like configuration with unit topological charge as a function of the applied magnetic field.}
  \label{fig:rho52:inst}
\end{figure}

In the previous sections we have studied the CME using the gluon fields of the quantum vacuum.
Such gluonic fields contain ultraviolet fluctuations both of perturbative and nonperturbative nature.
In particular, the topological charge density is subjected to strong fluctuations with quite
nontrivial localization properties, so that the CME should be realized locally in a vicinity of each topological lump.
In order to illustrate the existence of the CME in a more theoretical fashion, we have prepared a
smooth instanton-like gluon field with a unit topological charge on the lattice $14^4$ and applied
the external magnetic field to this configuration. In Figure~\ref{fig:rho52:inst} we show the behavior of the chirality
fluctuations and the fluctuations of longitudinal electric current at this instanton-like background.

The first indication of the existence of the CME comes from the observation of the increase of chirality fluctuations with increase the strength of the external magnetic field. Figure~\ref{fig:rho52:inst} illustrates that the external magnetic field, on average, enhances a local imbalance between approximately left and approximately right fermionic modes in the background of this $Q=1$ configuration. The effect may readily be understood as a consequence of the CME: the magnetic field tend to localize ``left'' and ``right'' fermionic modes in different parts of the space-time because the left and right modes have opposite magnetic moments. The external magnetic field acts as a chemical potential which provides a different weight to these modes. The left and right modes have opposite direction of the momentum with respect to the magnetic moment so that the external magnetic field tries to separate them spatially as well. Thus, the positive and negative chiralities do not compensate themselves in $\rho_5$, leading to increase of the chirality fluctuations.

The second signature in favor of the CME appears in the properties of the (squared) longitudinal electric current.  Notice that the``longitudinal'' current contains, in fact, contribution from the two components, $j_3$ and $j_0$, because they are indistinguishable from a geometrical point of view (we remind that we are working at zero temperature in external magnetic field $B = F_{12}$).
Therefore we also plot in Figure~\ref{fig:rho52:inst} the absolute value of the longitudinal current squared, $j_\parallel^2 = j_0^2 + j_3^2$ which turns out to be an increasing function of the magnetic field. This behavior is consistent with our understanding of the CME.

\begin{figure}
  \includegraphics[width=7cm,angle=0]{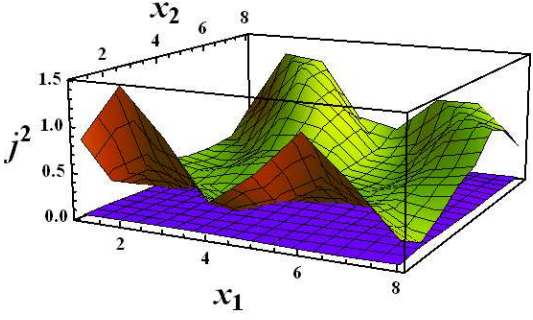}\\
  \caption{(color online). The squared components of the electric current (in arbitrary units) in a $12$-plane on the
  background of an instanton-like configuration with topological charge $Q=1$. The upper (green) sheet represents the
  spatial distribution of the longitudinal current $j^2_\parallel = j_0^2 + j_3^2$, and the lower (violet) sheet
  corresponds to the transverse current $j^2_\perp = j_1^2 + j_2^2$.}
  \label{fig:j2:3d:inst1}
\end{figure}

\begin{figure}
  \includegraphics[width=7cm,angle=0]{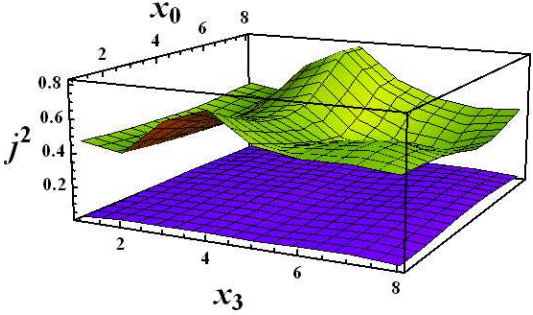}\\
  \caption{(color online). The same as in Figure~\ref{fig:j2:3d:inst1} but for a $03$-plane of the same gluon field.}
  \label{fig:j2:3d:inst2}
\end{figure}

In order to provide an illustration of the effect, we visualize the squared transverse and longitudinal electric current densities for the $Q=1$
configuration in the external magnetic field.
In Figures~\ref{fig:j2:3d:inst1} and \ref{fig:j2:3d:inst2} we show the current densities in the $12$ and $34$ cross-sections, respectively. One can clearly see that in a strong magnetic field the longitudinal component is always larger than the transverse one in agreement with the prediction of the CME.

Concluding this subsection we summarize, that our study of an instanton-like configuration of the gluonic field with the topological charge $Q=1$ supports the existence of the Chiral Magnetic Effect.

\section{Charge asymmetry: comparison with experiment}
\label{sec:fit}

Experimental signatures of the Chiral Magnetic Effect are usually discussed in terms of the following observables, suggested first in Ref.~\cite{Voloshin:04:1}:
\begin{eqnarray}
\label{observables}
a_{ab} = \frac{1}{N_{e}} \, \sum \limits_{e = 1}^{N_{e}} \, \frac{1}{N_{a} N_{b}} \,
\sum \limits_{i=1}^{N_{a}} \sum \limits_{j=1}^{N_{b}} \, \cos\lr{\phi_{i a} + \phi_{j b}},
\end{eqnarray}
where $a, b = \pm$ labels positively/negatively charged particles, $N_{e}$ is the total number of experimental events, $N_{a}$ and $N_{b}$ are the numbers of positively/negatively charged particles produced in each event, and $\phi_{i a}$, $\phi_{j b}$ are the azimuthal angles of the momenta of the produced particles w.r.t. the reaction plane.

These observables can be expressed in terms of the total charge separated in each event as: \cite{Kharzeev:08:1}:
\begin{eqnarray}
\label{observables_vs_delta}
a_{ab} = \frac{c \, \vev{\Delta_{a} \Delta_{b}}}{\vev{N_{a}} \vev{N_{b}}}\,,
\end{eqnarray}
where $a, b = \pm$, $\Delta_{+}$ is the difference of the total numbers of positively charged particles emitted above or below the reaction plane, $\Delta_{-}$ is the same quantity for negatively charged particles and $c$ is some coefficient (of order of unity) which depends on the characteristics of the hydrodynamical flow in the reaction. One also has $\vev{N_{a}} = \vev{N_{b}} = N_{q}$, where the average total number of the particles of the same charge per event $N_{q}$ (for Au-Au collisions) can be found, e.g., in Table 1 in \cite{Kharzeev:08:1}.

In order to compare experimental data with our results, it is convenient to consider the following combination of $a_{ab}$:
\begin{eqnarray}
\label{main_observable}
 a_{++} + a_{--} - 2 a_{+-} = \frac{\vev{\lr{\Delta Q}^{2}}}{N_{q}^{2}} = \frac{\vev{\lr{\Delta_{+} - \Delta_{-}}^{2}}}{N_{q}^{2}}\,,
\end{eqnarray}
where $\Delta Q = \Delta_{+} - \Delta_{-}$ - is the difference of total charges of particles emitted above and below the reaction plane. $\vev{\lr{\Delta Q}^{2}}$ can be estimated from our lattice data on $\vev{j_{\mu}^{2}}$ using the following rough model: consider a fireball of radius $R$ created during the heavy-ion collision, and assume that it emits positively or negatively charged particles from each element of surface $\delta\vec{\sigma}_{i}$, with $|\delta\vec{\sigma}_{i}| \sim \rho^{2}$. The quantity $\rho$ is some characteristic correlation length, so that emission of particles on each surface element of the area $\rho^2$ is statistically independent. $\Delta Q$ is calculated as the difference of the total charges emitted in some characteristic collision time $\tau$ from the upper and from the lower hemispheres of the surface of the fireball:
\begin{eqnarray}
\label{dQ_curr}
\delta Q = \tau \, \sum \limits_{i} \, \sign\lr{\delta\vec{\sigma}_{i} \cdot \vec{B}} \, \vec{j}_{i} \cdot \delta\vec{\sigma}_{i}\,.
\end{eqnarray}

 For the expectation value $\vev{\lr{\Delta Q}^{2}}$ one then has:
\begin{eqnarray}
& & \label{dQ_curr2}
\vev{\lr{\Delta Q}^{2}} = \tau^{2}  \, \sum \limits_{i} \, \vev{ \lr{\vec{j}_{i} \cdot \delta\vec{\sigma}_{i}}^{2} }
 \\
& &  = \tau^{2} \rho^{2} \, \sum \limits_{i} \, \vev{ \lr{\vec{j}_{i} \cdot n_{i}}^{2} } |\delta\vec{\sigma}_{i}| = \tau^{2} \rho^{2} \, \int d\vec{n} \, \vev{\vec{j} \cdot \vec{n}}^{2}\,.
\nonumber
\end{eqnarray}
We further assume that emission in different directions is independent: $\vev{j_{\mu} j_{\nu}} = 0$ for $\mu \neq \nu$. With these assumptions, we finally arrive at the following expression for $\vev{\lr{\Delta Q}^{2}}$:
\begin{eqnarray}
\label{dQ_curr2_fin}
\frac{\vev{\lr{\Delta Q}^{2}}}{N_{q}^{2}} = \frac{4 \pi \tau^{2} \rho^{2} R^{2}}{3 N_{q}^{2}}\, \lr{\vev{j_{\parallel}^{2}} + 2 \vev{j_{\perp}^{2}}}
\end{eqnarray}
where $j_{\parallel}$ and $j_{\perp}$ are, respectively, the currents along the magnetic field and perpendicular to it. We take the values of $N_{q}$ from Table 1 in \cite{Kharzeev:08:1} and use our lattice data on $\vev{j_{\parallel}^{2}}$ and $\vev{j_{\perp}^{2}}$ at $T = 1.12 T_c$. Here we take into account the fact that the expectation value of the squared spatial currents in the Minkowski and Euclidean spaces are the same.

\begin{figure}
  \includegraphics[width=6cm, angle=-90]{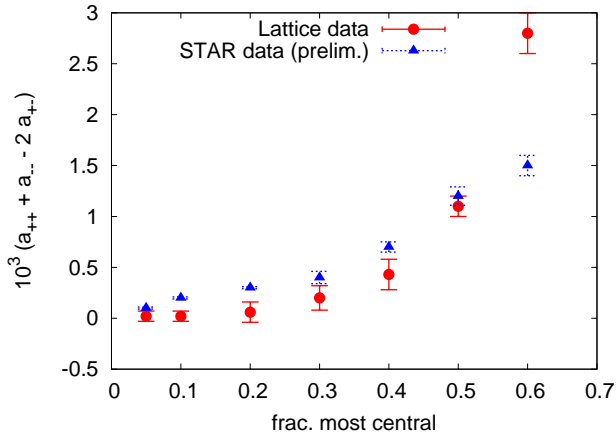}\\
  \caption{(color online). The average fluctuations of the difference of total electric charges of particles emitted above and below the reaction plane:
  comparison of the experimental observable of~\cite{Caines:09:1, Voloshin:08:1} (the filled triangles)
  with our lattice data (the filled circles).}
  \label{fig:cme_asymm_cmp}
\end{figure}

In Table 1 in \cite{Kharzeev:08:1} one can also find a relation between centrality (``\% most central'') and the ratio $b/R$, where $b$ is the impact parameter. From expression (A.12) in \cite{Kharzeev:08:1} one can roughly estimate the magnetic field during the collision as $e B \sim (0.1 b/R) \, \mbox{GeV}^{2}$. For simplicity, we assume that the magnetic field is nonzero and constant only during the time $\tau$. In order to obtain $\vev{\lr{\Delta Q}^{2}}$ at small values of magnetic field  we have used polynomial interpolation. The resulting plot is shown in Figure~\ref{fig:cme_asymm_cmp}.
Experimental points obtained by STAR collaboration are taken from the plots presented in~\cite{Caines:09:1, Voloshin:08:1}. The coefficient $\tau^{2} \rho^{2} R^{2}$ in (\ref{dQ_curr2_fin}) was fitted so that the lattice data and the theoretical data match best. One can take, for example, $\tau \sim 1 \, \mathrm{fm}$, $R \sim 5 \, \mathrm{fm}$, $\rho \sim 0.2 \, \mathrm{fm}$, which are reasonable values by order of magnitude, especially for such a rough model.

We observe a good agreement between our data and the experimental results. {\it Despite the agreement is seen at the quantitative level, the approximations used in our lattice technique suggest that this result should be considered as a qualitative indication rather than a solid proof}.
It should be noted that at $T = 1.12 T_c$ the dependence of $\vev{j^{2}}$ and hence of $\vev{\lr{\Delta Q}^{2}}$ on the magnetic field is rather weak, and, moreover, the current does not increase with magnetic field. We have to conclude that the dominant effect which determines the shape of the plot on (\ref{fig:cme_asymm_cmp}) is the decrease of $N_{q}$ with centrality (see Table 1 in \cite{Kharzeev:08:1}). It could be therefore much more informative to consider the observables $b_{ab}$ introduced in \cite{Kharzeev:08:1}, which do not contain the normalization factors $\frac{1}{N_{a} N_{b}}$. These observables can be directly compared with our data for the observable $\vev{j_{\mu}^{2}}$, both in the deconfinement phase and at zero temperature. Note also that since at zero temperature the current increases with magnetic field much faster than at high temperature, it might be not unreasonable to search for strong magnetic effects also in cold (and dense) nuclear matter.

\section{Conclusions}

In this paper we studied possible signatures of the Chiral Magnetic Effect~\cite{Kharzeev:98:1, Kharzeev:06:1, Kharzeev:08:1, Kharzeev:08:2} in quenched $SU\lr{2}$ lattice gauge theory using the chirally invariant Dirac operator. We observed the expected enhancement of the electric current in the direction of the magnetic field both in quantum configurations of the gluon fields and in a fixed instanton-like gauge fields
with nonzero topological charge. The CME at the instanton-like configurations is much more pronounced compared to the real
gluonic vacuum (an illustration can be found in Figures~\ref{fig:j2:3d:inst1} and~\ref{fig:j2:3d:inst2}).
The CME is based on the imbalance between left and right fermionic modes. The measure of the local imbalance is the chiral density of the quarks. On average the global chiral charge is zero, while the local fluctuations of the chiral density can be quite strong. We found (Figure~\ref{fig:rho52_vs_H}) that at zero temperature the chiral fluctuations are drastically (by two orders of magnitude) enhanced in the applied magnetic field.

As the temperature increases the chirality fluctuations increase and, simultaneously, they become less sensitive to the strength of the magnetic field. In the deconfinement phase the local chirality fluctuates much stronger compared to the fluctuations at zero temperature while the fluctuations themselves are practically independent on the strength of the magnetic field. The enhancement of the chirality fluctuations due to magnetic field can be illustrated by the following example: in strong magnetic field $q B = 2.5\,\mbox{GeV}^2$ at zero temperature $T=0$ the chirality fluctuations take the same value $\langle \rho_5^2\rangle \approx (220 \, \mbox{MeV})^6$ as in the case of zero magnetic field, $q B = 0$ and high temperature, $T = 1.12\, T_c$.

In the confinement and in the deconfinement phases the properties of the induced electric currents differ strongly. At zero temperature the longitudinal (i.e., directed along the magnetic field) currents grow with the increase of the strength of the magnetic field. This effect is accompanied by a (weaker) enhancement of the transverse currents, which we attribute to a transverse squeezing of the Landau levels (Figure~\ref{fig:j2vsH:T0}).

At nonzero temperature but still below the transition, $T=0.82 T_c$, the slopes of enhancement of both longitudinal and transverse electric currents become smaller. However, the longitudinal current is enhanced stronger than the transverse current (Figure~\ref{fig:j2vsH:T082}).

In the deconfinement phase at $T=1.12 T_c$ the fluctuations of the longitudinal currents are almost independent on the magnetic field. Moreover, the transverse components of the electric current decrease slightly as the magnetic field gets stronger. The fluctuations of the electric charge density become drastically suppressed in the deconfinement phase in the strong enough magnetic field (Figure~\ref{fig:j2vsH:T112}).

We also found that the presence of a nonzero {\it global} topological charge in the real gluonic configurations leads to a little effect on the longitudinal currents. This happens because the influence of a global charge is largely overshadowed by strong {\it local} fluctuations of the topological charge. We conclude, that in our simulations the CME is realized at local lumps of topological charge in the gluonic configuration.

We would like to make here a cautionary notice that our investigation utilized many approximations. We used two colors instead of three, our simulations are done in the quenched approximation, the fermionic propagators were truncated, and the system was studied in a finite volume. We found that the systematic errors associated with truncation, finite-size and finite-volume effects are smaller than our statistical errors, and we expect on general grounds that the reduced number of colors should not affect our results drastically. However the absence of the dynamical fermions may in general correct our results. We got a qualitative indication of the existence of the CME, while the quantitative features of the CME may be accessible only in simulations with dynamical quarks.

Finally, in real heavy ion collisions the magnetic field is perpendicular to the reaction plane. Consequently, the CME implies an imbalance between total electric charge observed above and below the reaction plane which is indeed seen in preliminary data published by STAR collaboration at RHIC~\cite{Caines:09:1, Voloshin:08:1}. Using a simple model of a fireball we obtain a good agreement between our data and experimental results of STAR Collaboration (Figure~\ref{fig:cme_asymm_cmp}) on the average fluctuations of the difference of total charges of particles emitted above and below the reaction plane.

\begin{acknowledgments}
The authors are grateful to V.G.~Bornyakov, A.S.~Gorsky, B.O.~Kerbikov, D.~Kharzeev, S.M.~Morozov, M.~M\"uller-Preussker, V.A.~Novikov, M.I.~Vysotsky, H.~Warringa and V.I.~Zakharov for interesting discussions.
This work was partly supported by Grants RFBR Nos. 08-02-00661-a, 06-02-17012, and DFG-RFBR 436 RUS, BRFBR F08D-005, a grant for scientific schools
No. NSh-679.2008.2, by the Federal Program of the Russian Ministry of Industry, Science and Technology No. 40.052.1.1.1112, by the Russian Federal Agency for Nuclear Power.  P.~V.~Buividovich is also supported by the Euler scholarship of the DAAD and by a scholarship of the Dynasty Foundation. The calculations were partially done on the MVS 50K at Moscow Joint Supercomputer Center.
\end{acknowledgments}


\begin{thebibliography}{26}
\bibitem{foot2}
Note that the words ``exact and Euclidean'' are very important here.
The exact zero modes in the Minkowski space or, as we will see below --
the near-zero modes in the Euclidean space are important for the CME.
\bibitem{foot1}
Notice that in a quenched theory the electromagnetic Debye screening is less effective compared to
a theory with virtual quarks because of the absence of the quark pair creation in the quenched case.
\end{thebibliography}
\end{document}